\documentclass[aps,prd,twocolumn,showpacs,preprintnumbers,superscriptaddress,nofootinbib]{revtex4-2}
\usepackage{float}
\usepackage{graphicx}
\usepackage{amsmath}
\usepackage{amsfonts}
\usepackage{amssymb}
\usepackage[bookmarks=true,bookmarksopen=true,
bookmarksnumbered=true,
colorlinks=true,linkcolor=black,
citecolor=red]{hyperref}
\usepackage{subfigure}
\usepackage{slashed}
\usepackage{setspace} 
\usepackage{ulem}
\usepackage{siunitx}   
\usepackage{braket}
\usepackage{url}
\usepackage{caption}
\usepackage{color}
\allowdisplaybreaks
\usepackage{enumerate}
\usepackage{graphicx}
\usepackage{dcolumn}
\usepackage{bm}

\begin{document}
	
	\preprint{APS/123-QED}

\title{{Lepton flavor violation in photon-induced electron-muon pairs at the HL-LHC}}

\author{\small M. A. Arroyo-Ureña }
\email{marco.arroyo@fcfm.buap.mx}
\affiliation{\small Facultad de Ciencias F\'isico-Matem\'aticas}
\affiliation{\small Centro Interdisciplinario de Investigaci\'on y Ense\~nanza de la Ciencia (CIIEC), Benem\'erita Universidad Aut\'onoma de Puebla, C.P. 72570, Puebla, Pue., M\'exico,}

\author{\small O. Félix-Beltrán }
\email{olga.felix@correo.buap.mx}
\affiliation{\small Facultad de Ciencias de la Electrónica, Benem\'erita Universidad Aut\'onoma de Puebla, Apartado Postal 1152, C.P. 72570, Puebla, Pue., M\'exico.}

\author{\small J. Hernández-Sánchez }
\email{jaime.hernandez@correo.buap.mx}
\affiliation{\small Facultad de Ciencias de la Electrónica, Benem\'erita Universidad Aut\'onoma de Puebla, Apartado Postal 1152, C.P. 72570, Puebla, Pue., M\'exico.}

\author{\small C. G. Honorato}
\email{carlosg.honorato@correo.buap.mx}
\affiliation{\small Facultad de Ciencias de la Electrónica, Benem\'erita Universidad Aut\'onoma de Puebla, Apartado Postal 1152, C.P. 72570, Puebla, Pue., M\'exico.}

\author{S. Rosado-Navarro}
\email{sebastian.rosado@protonmail.com}
\affiliation{Centro Interdisciplinario de Investigaci\'on y Ense\~nanza de la Ciencia (CIIEC), Benem\'erita Universidad Aut\'onoma de Puebla, C.P. 72570, Puebla, M\'exico.}

\begin{abstract}
 

We show the outstanding potential of the High-Luminosity LHC (HL-LHC) to discover charged lepton flavor violation (cLFV) via the ultra-peripheral process $\gamma\gamma \to e^\pm\mu^\mp$. Using a gauge-invariant Effective Field Theory (EFT) framework—consistent with the most stringent bounds from radiative decays—we perform a full Monte Carlo analysis with multivariate techniques. Our results show that a $5\sigma$ discovery is achievable with an integrated luminosity of $\mathcal{L}_{\rm int} \gtrsim 2700$ fb$^{-1}$ for favorable benchmark scenarios. This study establishes photon-fusion at the HL-LHC as a powerful probe of cLFV, strongly motivating dedicated searches by the experimental collaborations.

\end{abstract}

\keywords{EFT, LFV, HL-LHC.}
\maketitle

\section{Introduction}

In the Standard Model (SM), LFV is strictly forbidden. However, the observation of neutrino oscillations has confirmed that such particles have non-zero masses, which introduces LFV in the neutral sector. Within the SM framework, cLFV processes remain highly suppressed due to the GIM mechanism. Consequently, any experimental observation of a cLFV process would constitute unambiguous evidence for physics beyond the SM (BSM). In this work, we aim to explore the $\gamma\gamma e\mu$ interaction within an EFT framework and assess its prospects for experimental scrutiny at the HL-LHC~\cite{CERN-HL-LHC}. An analogous method, but in $e^-e^+$ colliders, is studied in~\cite{Arroyo-Urena:2025dxa}. This strategy leverages photon-photon interactions in ultraperipheral collisions (UPCs), where these type of interactions provide a pristine environment for studying electromagnetic phenomena, as they are precisely described by Quantum Electrodynamics (QED), and can be efficiently modeled using the equivalent photon approximation (EPA).

A confrontation of the EFT predictions against current experimental bounds, shows that the most stringent constraint on the diphoton operators originates from the loop-induced radiative decays $\mu \to e \gamma$~\cite{MEG:2016leq}. However, even after accommodating this constraint by suppressing the relevant theory parameters, a viable region of parameter space is found, providing strong motivation for experimental searches.


\section{Theoretical framework}\label{Theory}

The local cLFV $\gamma\gamma \bar{\ell}_i \ell_j$ interaction is described by the effective Lagrangian~\cite{Bowman:1978kz}:
\begin{align}\label{Lagrangian_Q}
		\mathcal{L}_{\rm eff} &= \left( G_{SR}^{ij}\, \bar{\ell}_{L_i}\ell_{R_j} + G_{SL}^{ij}\, \bar{\ell}_{R_i}\ell_{L_j} \right) F_{\mu\nu}F^{\mu\nu} \nonumber \\
		&+ \left( \tilde{G}_{SR}^{ij}\, \bar{\ell}_{L_i}\ell_{R_j} + \tilde{G}_{SL}^{ij}\, \bar{\ell}_{R_i}\ell_{L_j} \right) \tilde{F}_{\mu\nu}F^{\mu\nu} + \text{H.c.}\,,
	\end{align}
where $\ell_i,\,\ell_j = e,\,\mu,\,\tau$ denote the lepton flavors, $F_{\mu\nu}$ is the electromagnetic field strength tensor, and $\tilde{F}_{\mu\nu} = \frac{1}{2}\epsilon_{\mu\nu\sigma\lambda}F^{\sigma\lambda}$ is its dual. The couplings $G_{SR}^{ij}$, $G_{SL}^{ij}$, $\tilde{G}_{SR}^{ij}$, and $\tilde{G}_{SL}^{ij}$ are dimensionful, scaling as $\Lambda^{-3}$, where $\Lambda$ is the energy cutoff of the EFT. This scaling is characteristic of dimension-7 operators in an EFT extension of the SM, where LFV processes are forbidden at tree level but can be induced in scenarios BSM. 

The free effective coupling that has a direct impact on the process $\gamma\gamma e\mu$ is controlled by $|G_{e\mu}|$. This effective coupling encapsulates the relevant operator coefficients, and can be written as:
\begin{equation}\label{nonchiralcoupling}
	|G_{e\mu}|^2=2\big(|G_P^{e\mu}|^2+|G_S^{e\mu}|^2+|\tilde{G}_P^{e\mu}|^2+|\tilde{G}_S^{e\mu}|^2\big),
\end{equation}
where
\begin{align}\label{couplings_SP}
	G_S^{e\mu} &= \frac{G_{SR}^{e\mu} + G_{SL}^{e\mu}}{2}, \quad G_P^{e\mu} = \frac{G_{SR}^{e\mu} - G_{SL}^{e\mu}}{2}, \nonumber \\
	\tilde{G}_S^{e\mu} &= \frac{\tilde{G}_{SR}^{e\mu} + \tilde{G}_{SL}^{e\mu}}{2}, \quad \tilde{G}_P^{e\mu} = \frac{\tilde{G}_{SR}^{e\mu} - \tilde{G}_{SL}^{e\mu}}{2}\,.
\end{align}

\begin{figure}[!htb]
	\centering
	\includegraphics[scale=0.2]{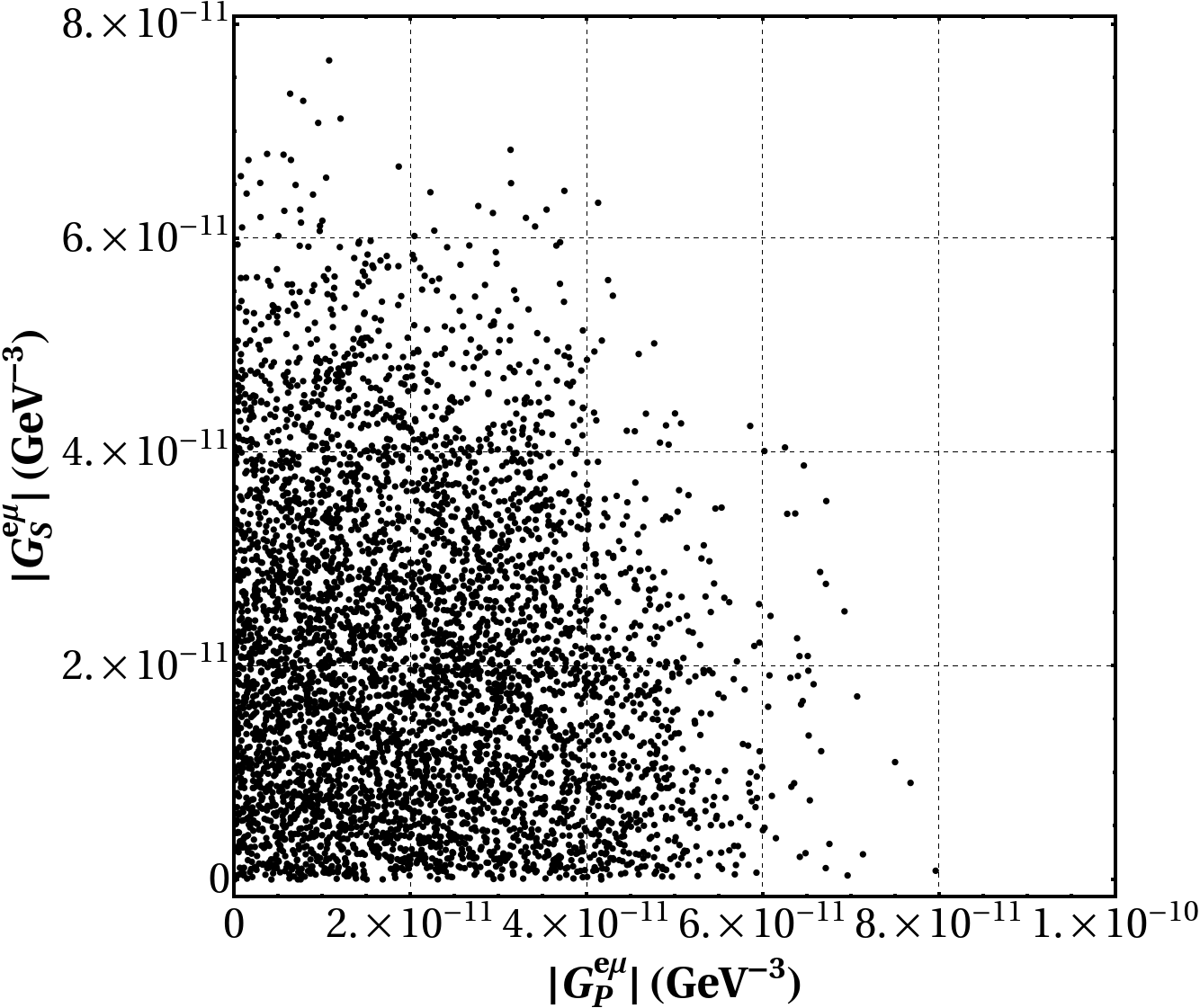}
	\caption{Parameter space in the $|G_S^{e\mu}| – |G_P^{e\mu}|$. We scan the intervals $|G_{S,\,P}^{e\mu}|, |\tilde{G}_{S,\,P}^{e\mu}|\in[0,10^{-9}]$ GeV$^{-3}$ and $\Lambda\in(100,\,5000)$ GeV. The black points represent those allowed by the constraints in Eqs.~\eqref{ul_Gij} and \eqref{nonchiralcoupling}.
	} \label{ParamSpace}
\end{figure}

From the upper limit on ${\rm BR}(\mu \to e \gamma)$, 
a stringent constraint on the effective coupling $|G_{e\mu}|$ is found~\cite{Fortuna:2023paj}:
\begin{align}\label{ul_Gij}
	|G_{\mu e}|&\lesssim 1.2\times10^{-10}\left(1+0.15\ln{\frac{\Lambda}{100~{\rm GeV}}}\right)^{-1}~{\rm GeV}^{-3}.\nonumber\\
\end{align} 
From Eqs.~\eqref{ul_Gij} and \eqref{nonchiralcoupling}, we obtain the parameter space presented in Fig.~\ref{ParamSpace}.   

For this study, we select the benchmark values of the cut-off scale such as $\Lambda \gg \sqrt{s_{\gamma\gamma}}$, where $\sqrt{s_{\gamma\gamma}}$ is the center-of-mass energy of the $\gamma\gamma$ system . This choice is motivated by the requirement that the EFT expansion remains consistent; specifically, for $\Lambda \gtrsim 2\sqrt{s_{\gamma\gamma}}$, contributions from a dimension-$d$ operator are suppressed by at least $(\sqrt{s_{\gamma\gamma}}/\Lambda)^{d-4}$, ensuring non-renormalizable terms remain under control. The center-of-mass energy required to produce an $e\mu$ pair is $\sqrt{s_{\gamma\gamma}} \gtrsim m_{\mu}$, which might naively suggest that a low effective scale $\Lambda$ could suffice. However, values of $\Lambda \lesssim 100\ \text{GeV}$ are incompatible with the EFT framework, as they would invalidate its core assumption that $\Lambda$ is significantly larger than the energy scale of the process. In fact, such a low $\Lambda$ would imply the existence of new particles with masses near or below the electroweak scale, which would require their explicit inclusion in the SM rather than a description through effective operators.

 Accordingly, we will explore three BP's to span the viable parameter space:
\begin{itemize}	
    \item {\bf BP1}: $\Lambda = \SI{1048}{\GeV}$,
    \begin{align*}
|G_S^{e\mu}| &= 3.85 \times 10^{-11}\,\text{GeV}^{-3},\, |G_P^{e\mu}| = 4.25 \times 10^{-11}\, \text{GeV}^{-3}, \\
|\tilde{G}_S^{e\mu}| &= 1.91 \times 10^{-12}\, \text{GeV}^{-3},\, |\tilde{G}_P^{e\mu}| = 2.42 \times 10^{-12}\, \text{GeV}^{-3}, \\     	  
    \end{align*}
        \item {\bf BP2}: $\Lambda = \SI{730}{\GeV}$,
       \begin{align*}
      	|G_S^{e\mu}| &= 5.22 \times 10^{-11}\, \text{GeV}^{-3},\, |G_P^{e\mu}| = 1.46 \times 10^{-11}\,\text{GeV}^{-3}, \\
       	|\tilde{G}_S^{e\mu}| &= 2.47 \times 10^{-11}\,\text{GeV}^{-3},\,|\tilde{G}_P^{e\mu}| = 5.96 \times 10^{-12}\,\text{GeV}^{-3},\\
        \end{align*}
          \item {\bf BP3}: $\Lambda = \SI{1859}{\GeV}$,
          \begin{align*}
 	|G_S^{e\mu}| &= 2.70 \times 10^{-11}\,\text{GeV}^{-3},\, |G_P^{e\mu}| = 1.92 \times 10^{-11}\, \text{GeV}^{-3},\\
 	|\tilde{G}_S^{e\mu}| &= 3.26 \times 10^{-11}\, \text{GeV}^{-3},\,	|\tilde{G}_P^{e\mu}| = 3.16 \times 10^{-11}\, \text{GeV}^{-3}.\\ 
          \end{align*}
\end{itemize}
    In addition to these values strictly satisfying stringent current experimental limits, the chosen cutoff scales ($\Lambda\sim 1-2$ TeV) are phenomenologically motivated, as they are consistent with interpreting the reported excesses near 95, 150, and 650 GeV as new physics resonances \cite{Bhattacharya:2025rfr,CMS:2023boe}. If one estimates $M_{\phi_k}\sim C_{e\mu}^k|G_{e\mu}|\Lambda^4$ ($k=1,2,3$), for our benchmarks, this connection requires coefficients $C_{e\mu}^k$ of $\mathcal{O}(1)$, suggesting a potential common UV origin that our cLFV search uniquely probes\footnote{Given that $|G_{e\mu}|$ has dimensions of GeV$^{-3}$, the mass $M_{\phi_k}$ have dimensions of GeV. }.

\section{Collider analysis}\label{Collider}

\subsection{Signal and background}

\begin{itemize}
	\item \textbf{Signal:} We are interested in searching for the final state of an opposite-sign $e\mu$ pair, which is produced via UPCs ($\gamma\gamma \to e\mu$) in proton-proton ($pp$) collisions.
		\item \textbf{Background:} The dominant SM backgrounds are $qq\to W^-W^+$, $\gamma\gamma\to W^+ W^-$, and $\gamma\gamma\to \tau^-\tau^+\to e\mu+\nu's$, with smaller contributions from misidentified hadronic processes. To mitigate the substantial background from pile-up, we employ the delphes\_card\_HLLHC-Pile-up.tcl\footnote{This card is made available upon request.} configuration. Our simulation includes pile-up with an average of 140 interactions per crossing, corresponding to HL-LHC conditions, via the \texttt{PileUpMerger} module and the pileup
per particle identification (PUPPI) algorithm. The sensitivity could be further improved beyond our conservative estimates by using forward proton spectrometers (roman pots) to tag exclusive events and suppress pile-up backgrounds~\cite{ATLAS:2024ztz, CMS:2024qjo}.

\end{itemize}


		The numerical cross sections for the signal and dominant background processes are presented in Tables~\ref{tb:XS-signal} and \ref{tb:XS-BGD}, respectively. These results incorporate the fiducial cuts $p_T^{(e,\,\mu)} > 10$ GeV, $|\eta^{(e,\,\mu)}| < 2.5$, $\Delta R(e\mu) > 0.4$.

\begin{table}[b!]
	\centering
	\caption{Production cross sections of the signal for the three BP's. The number of events is evaluated for $\mathcal{L}_{\rm int}=3$ ab$^{-1}$.}
	\label{tb:XS-signal}
	
	\renewcommand{\arraystretch}{1.2} 
	\setlength{\tabcolsep}{6pt} 
	
	\begin{tabular}{|c|c|c|}
		\hline
		BP's &  $\sigma(\gamma\gamma\to e\mu)$(fb)& Events \\
		\hline 
		\hline
		BP1 & $0.04$ & 120\\
		\hline
		BP2 & $0.035$ & 105\\
		\hline
		BP3 & $0.013$ & 39\\
		\hline 
	\end{tabular}
\end{table}

\begin{table}[b!]
	\caption{Production cross sections of relevant SM background processes. The number of events is evaluated for $\mathcal{L}_{\rm int}=3$ ab$^{-1}$.}\label{tb:XS-BGD}
	\begin{centering}
		\begin{tabular}{|c|c|c|}
			\hline 
			Process & cross section (fb)& Events\tabularnewline
			\hline 
			\hline 
			$\gamma\gamma \to WW\to e\nu_e\mu \nu_\mu$ & $0.623$& 1869 \tabularnewline
			\hline
			$\gamma\gamma\to\tau\tau\to e\nu_e \nu_\tau\mu \nu_\mu\nu_\tau$ & $28.2$ & 84600\tabularnewline
			\hline
			$qq\to WW\to e\nu_e \mu \nu_\mu$ & $890$ & 2.67$\times 10^{6}$\tabularnewline
			\hline
		\end{tabular}
		\par\end{centering}
\end{table}


Signal and background events are generated with \texttt{MadGraph5} (\texttt{LanHEP/UFO}) -using the electric dipole form factor (EDFF) to model UPC's-, showered with \texttt{Pythia8}, and passed through \texttt{Delphes3}~\cite{Semenov:2014rea, Degrande:2011ua, Alwall_2011, Sjostrand:2014zea, deFavereau:2013fsa, DelphesHLLHCCard, Shao:2022cly}.

\subsection{Multivariate Analysis}
The kinematic analysis revealed that the individual observables had relatively weak discriminating power to separate the signal from background. To enhance the selection efficacy, we therefore employed a Multivariate Analysis (MVA) approach, which combines these observables into a single, more powerful discriminant. We use a Boosted Decision Tree (BDT) method~\cite{Coadou:2022nsh} based on the XGBoost library's gradient boosting implementation. The BDT classifiers were trained on a set of kinematic variables associated to the final-state objects, including the transverse momentum ($p_T$), pseudorapidity ($\eta$), angular separation ($\Delta R$), transverse and invariant masses ($M_{T}$, $M_{\rm inv}$), azimutal angle ($\phi$).

The BDT selection is optimized to maximize the figure of merit, $i.e$., the \textit{signal significance}, defined as $\mathcal{S}=N_S/\sqrt{N_S+N_B+(\kappa\cdot N_B)^2}$~\cite{Cowan2021}, where $N_S$ and $N_B$ are the number of signal and background candidates, respectively. The factor $\kappa=5\%$ stands for a realistic systematic uncertainty~\cite{ATLAS:2021pce}.
		
\begin{figure}[!htb]
	\centering
	\includegraphics[scale=0.18]{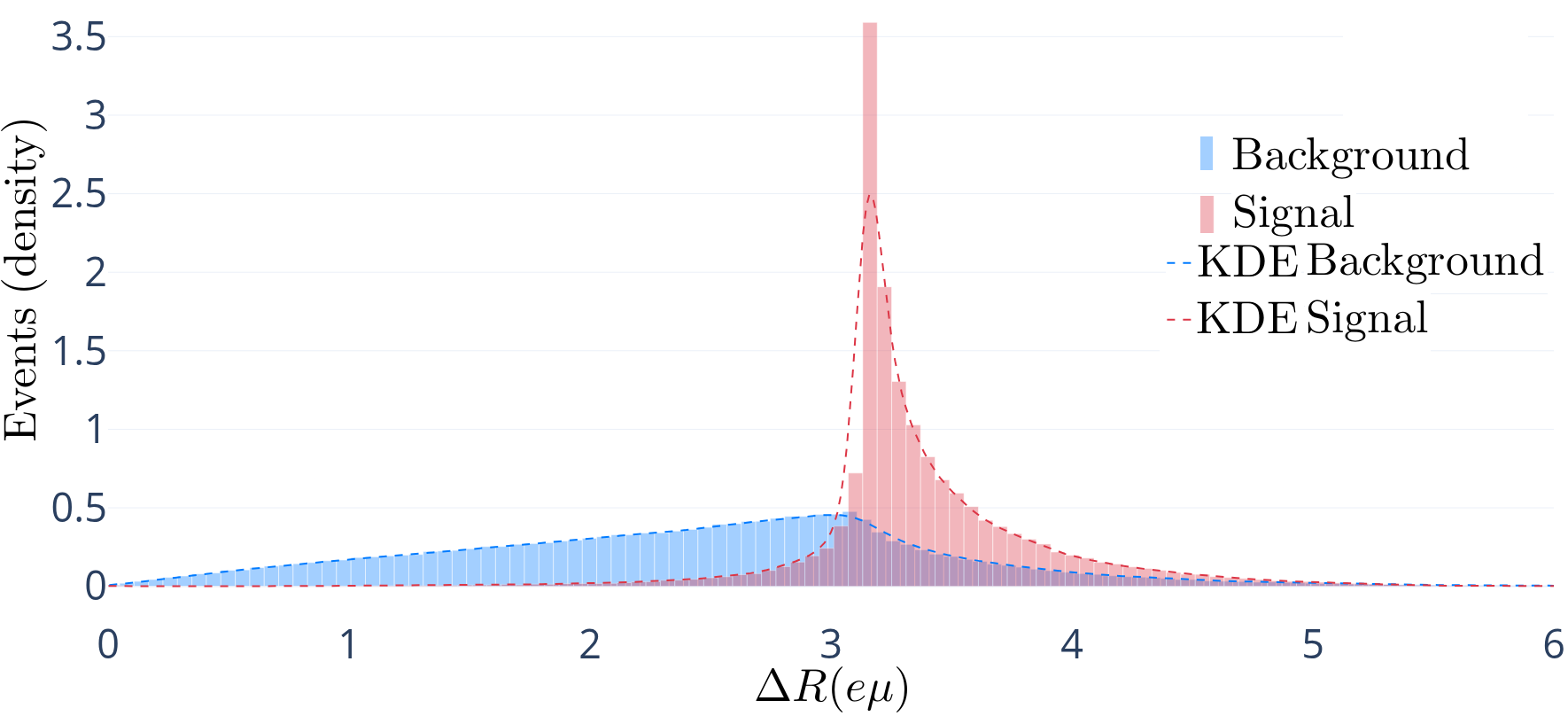}
	\caption{Angular separation $\Delta R(e\mu)$. We also include the fit Kernel Density Estimation (KDE).}  
	\label{distributions}
\end{figure}

Building upon the previous MVA, the most powerful discriminating observables for distinguishing the signal from background processes are: $i)$  the angular separation between the electron and muon $\Delta R(e\mu)$, $ii)$ transverse mass of the electron $M_T(e)=\sqrt{2p_T^e\slashed{E}_T(1-\cos\phi_{e\slashed{E}_T})}$, and $iii)$ missing energy transverse $\slashed{E}_T$. The distribution of the most significant variable, which exhibit a pronounced separation between the signal and the several background channels, is presented in Fig.~\ref{distributions}. 

On the other hand, the absence of over-training is confirmed by a Kolmogorov-Smirnov (KS) test, which yields high scores of 0.49 for background and 0.34 for signal.

\begin{figure}[!htb]
		\includegraphics[scale=0.23]{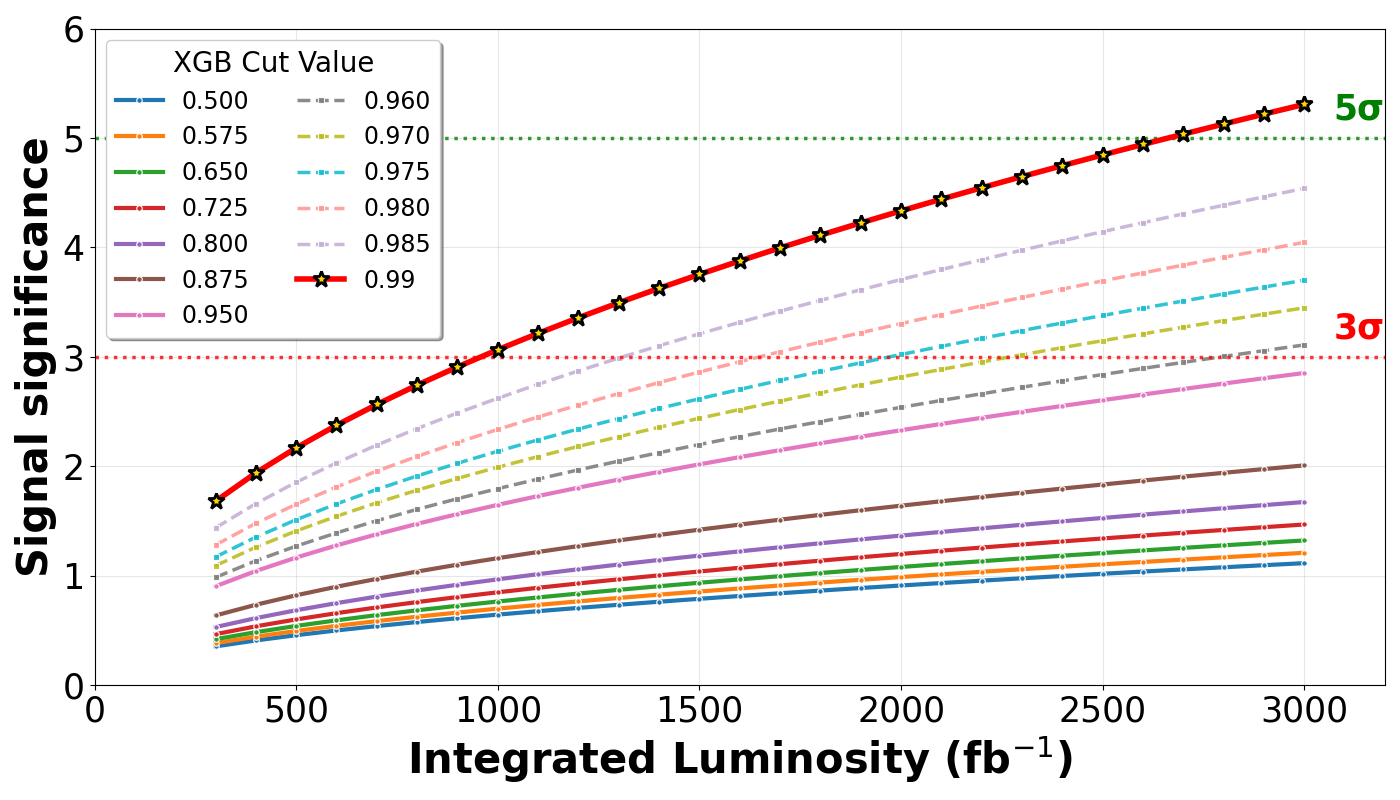}
		\caption{\textit{Signal significance} as a function of integrated luminosity and the cut on the BDT prediction for BP1. Systematic uncertainties ($\kappa =5\%$) are also considered.}  
		\label{fig:significance}
	\end{figure}

The exceptional discovery potential of the signal at the HL-LHC is demonstrated by our key results in Fig.~\ref{fig:significance}, which show the projected signal significance as a function of integrated luminosity and the corresponding BDT classifier outputs for BP1.

 The optimal scenario corresponds to BP1, we obtain a signal significance of $5\sigma$ for $\mathcal{L}_{\rm int} \gtrsim 2700\,{\rm fb}^{-1}$ and a BDT cut of $0.99$. Similar predictions are found for BP2, which predicts that a $5\sigma$ significance is achievable with $\mathcal{L}_{\rm int} \gtrsim 2800\,{\rm fb}^{-1}$. BP3 predicts up to approximately $2\sigma$ for an integrated luminosity of $\mathcal{L}_{\rm int} \approx 3000\,{\rm fb}^{-1}$. 

\section{Conclusions}\label{Conclusions}

We have shown the unique capability of the HL-LHC to probe charged lepton flavor violation through the ultra-peripheral process $pp\to p(\gamma\gamma)p\to p(e^\pm \mu^\mp)p$. Our analysis, based on a dimension-7 effective field theory framework, reveals a direct connection between the experimental sensitivity and the underlying new physics scale $\Lambda$. Crucially, the benchmark points BP1 ($\Lambda=1048$ GeV), BP2 ($\Lambda=730$ GeV), and BP3 ($\Lambda=1859$ GeV) span a phenomenologically meaningful range of cutoff scales. The fact that BP2—with the lowest allowed $\Lambda$-yields a strong signal highlights how this channel is particularly sensitive to new physics at the TeV scale. Meanwhile, BP3 demonstrates the challenge of probing higher scales, still remains accessible at high luminosities. Notably, a $5\sigma$ discovery is achievable for $\Lambda\sim 1$ TeV with $\mathcal{L}_{\rm int} \gtrsim 2700$ fb$^{-1}$, while scales up to $\Lambda\sim 2$ TeV may be probed at the approximately $2\sigma$ level with $\mathcal{L}_{\rm int} \approx 3000$ fb$^{-1}$. This clear correlation between the cutoff scale and the required luminosity provides valuable guidance for future searches.

These results strongly motivate dedicated experimental efforts to search for 
$\gamma\gamma\to e\mu$ in early HL-LHC data. Future colliders with higher energy reach will be essential to extend this probe to even higher scales, offering a powerful tool to explore lepton flavor violation beyond the TeV frontier~\cite{Cepeda:2019klc, Arkani-Hamed:2015vfh}.

The high photon fluxes in Pb-Pb collisions offer a distinct advantage: they can substantially enhance the production cross section. This enhancement compensates for the lower integrated luminosity, making it a highly promising channel. The net effect is an estimated increase in the number of observable events by approximately an order of magnitude compared to $pp$ collisions \cite{Klusek-Gawenda:2025ffh, DEnterria:2019cpw}.
\section*{Acknowledgments}

We are grateful to Fabiola Fortuna for her invaluable comments. This work was supported by the Estancias Posdoctorales por México program, the Sistema Nacional de Investigadores e Investigadoras (SNII), VIEP-BUAP, and PRODEP (México), and the SECIHTI project No. CBF-2025-G-1187.

\bibliography{references}

\begin{thebibliography}{23}%
\makeatletter
\providecommand \@ifxundefined [1]{%
 \@ifx{#1\undefined}
}%
\providecommand \@ifnum [1]{%
 \ifnum #1\expandafter \@firstoftwo
 \else \expandafter \@secondoftwo
 \fi
}%
\providecommand \@ifx [1]{%
 \ifx #1\expandafter \@firstoftwo
 \else \expandafter \@secondoftwo
 \fi
}%
\providecommand \natexlab [1]{#1}%
\providecommand \enquote  [1]{``#1''}%
\providecommand \bibnamefont  [1]{#1}%
\providecommand \bibfnamefont [1]{#1}%
\providecommand \citenamefont [1]{#1}%
\providecommand \href@noop [0]{\@secondoftwo}%
\providecommand \href [0]{\begingroup \@sanitize@url \@href}%
\providecommand \@href[1]{\@@startlink{#1}\@@href}%
\providecommand \@@href[1]{\endgroup#1\@@endlink}%
\providecommand \@sanitize@url [0]{\catcode `\\12\catcode `\$12\catcode
  `\&12\catcode `\#12\catcode `\^12\catcode `\_12\catcode `\%12\relax}%
\providecommand \@@startlink[1]{}%
\providecommand \@@endlink[0]{}%
\providecommand \url  [0]{\begingroup\@sanitize@url \@url }%
\providecommand \@url [1]{\endgroup\@href {#1}{\urlprefix }}%
\providecommand \urlprefix  [0]{URL }%
\providecommand \Eprint [0]{\href }%
\providecommand \doibase [0]{https://doi.org/}%
\providecommand \selectlanguage [0]{\@gobble}%
\providecommand \bibinfo  [0]{\@secondoftwo}%
\providecommand \bibfield  [0]{\@secondoftwo}%
\providecommand \translation [1]{[#1]}%
\providecommand \BibitemOpen [0]{}%
\providecommand \bibitemStop [0]{}%
\providecommand \bibitemNoStop [0]{.\EOS\space}%
\providecommand \EOS [0]{\spacefactor3000\relax}%
\providecommand \BibitemShut  [1]{\csname bibitem#1\endcsname}%
\let\auto@bib@innerbib\@empty
\bibitem [{\citenamefont {CERN}(2024)}]{CERN-HL-LHC}%
  \BibitemOpen
  \bibfield  {author} {\bibinfo {author} {\bibnamefont {CERN}},\ }\href@noop {}
  {\bibinfo {title} {High-luminosity lhc}},\ \bibinfo {howpublished}
  {\url{https://home.cern/science/accelerators/high-luminosity-lhc}} (\bibinfo
  {year} {2024}),\ \bibinfo {note} {accessed: 2024-05-16}\BibitemShut {NoStop}%
\bibitem [{\citenamefont {Arroyo-Ure{\~n}a}\ \emph {et~al.}(2025)\citenamefont
  {Arroyo-Ure{\~n}a}, \citenamefont {Gait{\'a}n}, \citenamefont {Mar{\'\i}n},
  \citenamefont {Salazar},\ and\ \citenamefont
  {Villanueva-Utrilla}}]{Arroyo-Urena:2025dxa}%
  \BibitemOpen
  \bibfield  {author} {\bibinfo {author} {\bibfnamefont {M.~A.}\ \bibnamefont
  {Arroyo-Ure{\~n}a}}, \bibinfo {author} {\bibfnamefont {R.}~\bibnamefont
  {Gait{\'a}n}}, \bibinfo {author} {\bibfnamefont {M.}~\bibnamefont
  {Mar{\'\i}n}}, \bibinfo {author} {\bibfnamefont {H.}~\bibnamefont
  {Salazar}},\ and\ \bibinfo {author} {\bibfnamefont {M.~G.}\ \bibnamefont
  {Villanueva-Utrilla}},\ }\bibfield  {title} {\bibinfo {title} {{Searching for
  the LFV $\gamma\gamma e\mu$ interaction at $e^- e^+$ colliders}},\
  }\href@noop {} {\  (\bibinfo {year} {2025})},\ \Eprint
  {https://arxiv.org/abs/2504.12431} {arXiv:2504.12431 [hep-ph]} \BibitemShut
  {NoStop}%
\bibitem [{\citenamefont {Baldini}\ \emph {et~al.}(2016)\citenamefont {Baldini}
  \emph {et~al.}}]{MEG:2016leq}%
  \BibitemOpen
  \bibfield  {author} {\bibinfo {author} {\bibfnamefont {A.~M.}\ \bibnamefont
  {Baldini}} \emph {et~al.} (\bibinfo {collaboration} {MEG}),\ }\bibfield
  {title} {\bibinfo {title} {{Search for the lepton flavour violating decay
  $\mu ^+ \rightarrow \mathrm {e}^+ \gamma $ with the full dataset of the MEG
  experiment}},\ }\href {https://doi.org/10.1140/epjc/s10052-016-4271-x}
  {\bibfield  {journal} {\bibinfo  {journal} {Eur. Phys. J. C}\ }\textbf
  {\bibinfo {volume} {76}},\ \bibinfo {pages} {434} (\bibinfo {year} {2016})},\
  \Eprint {https://arxiv.org/abs/1605.05081} {arXiv:1605.05081 [hep-ex]}
  \BibitemShut {NoStop}%
\bibitem [{\citenamefont {Bowman}\ \emph {et~al.}(1978)\citenamefont {Bowman},
  \citenamefont {Cheng}, \citenamefont {Li},\ and\ \citenamefont
  {Matis}}]{Bowman:1978kz}%
  \BibitemOpen
  \bibfield  {author} {\bibinfo {author} {\bibfnamefont {J.~D.}\ \bibnamefont
  {Bowman}}, \bibinfo {author} {\bibfnamefont {T.~P.}\ \bibnamefont {Cheng}},
  \bibinfo {author} {\bibfnamefont {L.-F.}\ \bibnamefont {Li}},\ and\ \bibinfo
  {author} {\bibfnamefont {H.~S.}\ \bibnamefont {Matis}},\ }\bibfield  {title}
  {\bibinfo {title} {{New Upper Limit for $\mu\to e\gamma\gamma$}},\ }\href
  {https://doi.org/10.1103/PhysRevLett.41.442} {\bibfield  {journal} {\bibinfo
  {journal} {Phys. Rev. Lett.}\ }\textbf {\bibinfo {volume} {41}},\ \bibinfo
  {pages} {442} (\bibinfo {year} {1978})}\BibitemShut {NoStop}%
\bibitem [{\citenamefont {Fortuna}\ \emph {et~al.}(2023)\citenamefont
  {Fortuna}, \citenamefont {Marcano}, \citenamefont {Mar\'\i{}n},\ and\
  \citenamefont {Roig}}]{Fortuna:2023paj}%
  \BibitemOpen
  \bibfield  {author} {\bibinfo {author} {\bibfnamefont {F.}~\bibnamefont
  {Fortuna}}, \bibinfo {author} {\bibfnamefont {X.}~\bibnamefont {Marcano}},
  \bibinfo {author} {\bibfnamefont {M.}~\bibnamefont {Mar\'\i{}n}},\ and\
  \bibinfo {author} {\bibfnamefont {P.}~\bibnamefont {Roig}},\ }\bibfield
  {title} {\bibinfo {title} {{Lepton flavor violation from diphoton effective
  interactions}},\ }\href {https://doi.org/10.1103/PhysRevD.108.015008}
  {\bibfield  {journal} {\bibinfo  {journal} {Phys. Rev. D}\ }\textbf {\bibinfo
  {volume} {108}},\ \bibinfo {pages} {015008} (\bibinfo {year} {2023})},\
  \Eprint {https://arxiv.org/abs/2305.04974} {arXiv:2305.04974 [hep-ph]}
  \BibitemShut {NoStop}%
\bibitem [{\citenamefont {Bhattacharya}\ \emph {et~al.}(2025)\citenamefont
  {Bhattacharya}, \citenamefont {Lieberman}, \citenamefont {Kumar},
  \citenamefont {Crivellin}, \citenamefont {Fang}, \citenamefont {Mazini},\
  and\ \citenamefont {Mellado}}]{Bhattacharya:2025rfr}%
  \BibitemOpen
  \bibfield  {author} {\bibinfo {author} {\bibfnamefont {S.}~\bibnamefont
  {Bhattacharya}}, \bibinfo {author} {\bibfnamefont {B.}~\bibnamefont
  {Lieberman}}, \bibinfo {author} {\bibfnamefont {M.}~\bibnamefont {Kumar}},
  \bibinfo {author} {\bibfnamefont {A.}~\bibnamefont {Crivellin}}, \bibinfo
  {author} {\bibfnamefont {Y.}~\bibnamefont {Fang}}, \bibinfo {author}
  {\bibfnamefont {R.}~\bibnamefont {Mazini}},\ and\ \bibinfo {author}
  {\bibfnamefont {B.}~\bibnamefont {Mellado}},\ }\bibfield  {title} {\bibinfo
  {title} {{Emerging Excess Consistent with a Narrow Resonance at 152 GeV in
  High-Energy Proton-Proton Collisions}},\ }\href@noop {} {\  (\bibinfo {year}
  {2025})},\ \Eprint {https://arxiv.org/abs/2503.16245} {arXiv:2503.16245
  [hep-ph]} \BibitemShut {NoStop}%
\bibitem [{\citenamefont {Tumasyan}\ \emph {et~al.}(2024)\citenamefont
  {Tumasyan} \emph {et~al.}}]{CMS:2023boe}%
  \BibitemOpen
  \bibfield  {author} {\bibinfo {author} {\bibfnamefont {A.}~\bibnamefont
  {Tumasyan}} \emph {et~al.} (\bibinfo {collaboration} {CMS}),\ }\bibfield
  {title} {\bibinfo {title} {{Search for a new resonance decaying into two
  spin-0 bosons in a final state with two photons and two bottom quarks in
  proton-proton collisions at $ \sqrt{s} $ = 13 TeV}},\ }\href
  {https://doi.org/10.1007/JHEP05(2024)316} {\bibfield  {journal} {\bibinfo
  {journal} {JHEP}\ }\textbf {\bibinfo {volume} {05}},\ \bibinfo {pages}
  {316}},\ \Eprint {https://arxiv.org/abs/2310.01643} {arXiv:2310.01643
  [hep-ex]} \BibitemShut {NoStop}%
\bibitem [{\citenamefont {Aad}\ \emph {et~al.}(2024)\citenamefont {Aad} \emph
  {et~al.}}]{ATLAS:2024ztz}%
  \BibitemOpen
  \bibfield  {author} {\bibinfo {author} {\bibfnamefont {G.}~\bibnamefont
  {Aad}} \emph {et~al.} (\bibinfo {collaboration} {ATLAS}),\ }\bibfield
  {title} {\bibinfo {title} {{Performance of the ATLAS forward proton
  Time-of-Flight detector in Run 2}},\ }\href
  {https://doi.org/10.1088/1748-0221/19/05/P05054} {\bibfield  {journal}
  {\bibinfo  {journal} {JINST}\ }\textbf {\bibinfo {volume} {19}}\bibfield
  {number} {\bibinfo  {number} { (05)},\ \bibinfo {pages} {P05054}},\ }\Eprint
  {https://arxiv.org/abs/2402.06438} {arXiv:2402.06438 [physics.ins-det]}
  \BibitemShut {NoStop}%
\bibitem [{\citenamefont {Hayrapetyan}\ \emph {et~al.}(2024)\citenamefont
  {Hayrapetyan} \emph {et~al.}}]{CMS:2024qjo}%
  \BibitemOpen
  \bibfield  {author} {\bibinfo {author} {\bibfnamefont {A.}~\bibnamefont
  {Hayrapetyan}} \emph {et~al.} (\bibinfo {collaboration} {CMS}),\ }\bibfield
  {title} {\bibinfo {title} {{Observation of $\gamma\gamma\to\tau\tau$ in
  proton-proton collisions and limits on the anomalous electromagnetic moments
  of the $\tau$ lepton}},\ }\href {https://doi.org/10.1088/1361-6633/ad6fcb}
  {\bibfield  {journal} {\bibinfo  {journal} {Rept. Prog. Phys.}\ }\textbf
  {\bibinfo {volume} {87}},\ \bibinfo {pages} {107801} (\bibinfo {year}
  {2024})},\ \Eprint {https://arxiv.org/abs/2406.03975} {arXiv:2406.03975
  [hep-ex]} \BibitemShut {NoStop}%
\bibitem [{\citenamefont {Semenov}(2016)}]{Semenov:2014rea}%
  \BibitemOpen
  \bibfield  {author} {\bibinfo {author} {\bibfnamefont {A.}~\bibnamefont
  {Semenov}},\ }\bibfield  {title} {\bibinfo {title} {{LanHEP \textemdash{} A
  package for automatic generation of Feynman rules from the Lagrangian.
  Version 3.2}},\ }\href {https://doi.org/10.1016/j.cpc.2016.01.003} {\bibfield
   {journal} {\bibinfo  {journal} {Comput. Phys. Commun.}\ }\textbf {\bibinfo
  {volume} {201}},\ \bibinfo {pages} {167} (\bibinfo {year} {2016})},\ \Eprint
  {https://arxiv.org/abs/1412.5016} {arXiv:1412.5016 [physics.comp-ph]}
  \BibitemShut {NoStop}%
\bibitem [{\citenamefont {Degrande}\ \emph {et~al.}(2012)\citenamefont
  {Degrande}, \citenamefont {Duhr}, \citenamefont {Fuks}, \citenamefont
  {Grellscheid}, \citenamefont {Mattelaer},\ and\ \citenamefont
  {Reiter}}]{Degrande:2011ua}%
  \BibitemOpen
  \bibfield  {author} {\bibinfo {author} {\bibfnamefont {C.}~\bibnamefont
  {Degrande}}, \bibinfo {author} {\bibfnamefont {C.}~\bibnamefont {Duhr}},
  \bibinfo {author} {\bibfnamefont {B.}~\bibnamefont {Fuks}}, \bibinfo {author}
  {\bibfnamefont {D.}~\bibnamefont {Grellscheid}}, \bibinfo {author}
  {\bibfnamefont {O.}~\bibnamefont {Mattelaer}},\ and\ \bibinfo {author}
  {\bibfnamefont {T.}~\bibnamefont {Reiter}},\ }\bibfield  {title} {\bibinfo
  {title} {{UFO - The Universal FeynRules Output}},\ }\href
  {https://doi.org/10.1016/j.cpc.2012.01.022} {\bibfield  {journal} {\bibinfo
  {journal} {Comput. Phys. Commun.}\ }\textbf {\bibinfo {volume} {183}},\
  \bibinfo {pages} {1201} (\bibinfo {year} {2012})},\ \Eprint
  {https://arxiv.org/abs/1108.2040} {arXiv:1108.2040 [hep-ph]} \BibitemShut
  {NoStop}%
\bibitem [{\citenamefont {Alwall}\ \emph {et~al.}(2011)\citenamefont {Alwall},
  \citenamefont {Herquet}, \citenamefont {Maltoni}, \citenamefont {Mattelaer},\
  and\ \citenamefont {Stelzer}}]{Alwall_2011}%
  \BibitemOpen
  \bibfield  {author} {\bibinfo {author} {\bibfnamefont {J.}~\bibnamefont
  {Alwall}}, \bibinfo {author} {\bibfnamefont {M.}~\bibnamefont {Herquet}},
  \bibinfo {author} {\bibfnamefont {F.}~\bibnamefont {Maltoni}}, \bibinfo
  {author} {\bibfnamefont {O.}~\bibnamefont {Mattelaer}},\ and\ \bibinfo
  {author} {\bibfnamefont {T.}~\bibnamefont {Stelzer}},\ }\bibfield  {title}
  {\bibinfo {title} {Madgraph 5: going beyond},\ }\bibfield  {journal}
  {\bibinfo  {journal} {Journal of High Energy Physics}\ }\textbf {\bibinfo
  {volume} {2011}},\ \href {https://doi.org/10.1007/jhep06(2011)128}
  {10.1007/jhep06(2011)128} (\bibinfo {year} {2011})\BibitemShut {NoStop}%
\bibitem [{\citenamefont {Sj\"ostrand}\ \emph {et~al.}(2015)\citenamefont
  {Sj\"ostrand}, \citenamefont {Ask}, \citenamefont {Christiansen},
  \citenamefont {Corke}, \citenamefont {Desai}, \citenamefont {Ilten},
  \citenamefont {Mrenna}, \citenamefont {Prestel}, \citenamefont {Rasmussen},\
  and\ \citenamefont {Skands}}]{Sjostrand:2014zea}%
  \BibitemOpen
  \bibfield  {author} {\bibinfo {author} {\bibfnamefont {T.}~\bibnamefont
  {Sj\"ostrand}}, \bibinfo {author} {\bibfnamefont {S.}~\bibnamefont {Ask}},
  \bibinfo {author} {\bibfnamefont {J.~R.}\ \bibnamefont {Christiansen}},
  \bibinfo {author} {\bibfnamefont {R.}~\bibnamefont {Corke}}, \bibinfo
  {author} {\bibfnamefont {N.}~\bibnamefont {Desai}}, \bibinfo {author}
  {\bibfnamefont {P.}~\bibnamefont {Ilten}}, \bibinfo {author} {\bibfnamefont
  {S.}~\bibnamefont {Mrenna}}, \bibinfo {author} {\bibfnamefont
  {S.}~\bibnamefont {Prestel}}, \bibinfo {author} {\bibfnamefont {C.~O.}\
  \bibnamefont {Rasmussen}},\ and\ \bibinfo {author} {\bibfnamefont {P.~Z.}\
  \bibnamefont {Skands}},\ }\bibfield  {title} {\bibinfo {title} {{An
  introduction to PYTHIA 8.2}},\ }\href
  {https://doi.org/10.1016/j.cpc.2015.01.024} {\bibfield  {journal} {\bibinfo
  {journal} {Comput. Phys. Commun.}\ }\textbf {\bibinfo {volume} {191}},\
  \bibinfo {pages} {159} (\bibinfo {year} {2015})},\ \Eprint
  {https://arxiv.org/abs/1410.3012} {arXiv:1410.3012 [hep-ph]} \BibitemShut
  {NoStop}%
\bibitem [{\citenamefont {de~Favereau}\ \emph {et~al.}(2014)\citenamefont
  {de~Favereau}, \citenamefont {Delaere}, \citenamefont {Demin}, \citenamefont
  {Giammanco}, \citenamefont {Lema\^\i{}tre}, \citenamefont {Mertens},\ and\
  \citenamefont {Selvaggi}}]{deFavereau:2013fsa}%
  \BibitemOpen
  \bibfield  {author} {\bibinfo {author} {\bibfnamefont {J.}~\bibnamefont
  {de~Favereau}}, \bibinfo {author} {\bibfnamefont {C.}~\bibnamefont
  {Delaere}}, \bibinfo {author} {\bibfnamefont {P.}~\bibnamefont {Demin}},
  \bibinfo {author} {\bibfnamefont {A.}~\bibnamefont {Giammanco}}, \bibinfo
  {author} {\bibfnamefont {V.}~\bibnamefont {Lema\^\i{}tre}}, \bibinfo {author}
  {\bibfnamefont {A.}~\bibnamefont {Mertens}},\ and\ \bibinfo {author}
  {\bibfnamefont {M.}~\bibnamefont {Selvaggi}} (\bibinfo {collaboration}
  {DELPHES 3}),\ }\bibfield  {title} {\bibinfo {title} {{DELPHES 3, A modular
  framework for fast simulation of a generic collider experiment}},\ }\href
  {https://doi.org/10.1007/JHEP02(2014)057} {\bibfield  {journal} {\bibinfo
  {journal} {JHEP}\ }\textbf {\bibinfo {volume} {02}},\ \bibinfo {pages}
  {057}},\ \Eprint {https://arxiv.org/abs/1307.6346} {arXiv:1307.6346 [hep-ex]}
  \BibitemShut {NoStop}%
\bibitem [{\citenamefont {Collaboration}(2022)}]{DelphesHLLHCCard}%
  \BibitemOpen
  \bibfield  {author} {\bibinfo {author} {\bibfnamefont {D.}~\bibnamefont
  {Collaboration}},\ }\href
  {https://github.com/delphes/delphes/blob/master/cards/delphes_card_HLLHC.tcl}
  {\emph {\bibinfo {title} {Delphes 3 HL-LHC Card}}} (\bibinfo {year} {2022}),\
  \bibinfo {note} {gitHub repository:
  \url{https://github.com/delphes/delphes}}\BibitemShut {NoStop}%
\bibitem [{\citenamefont {Shao}\ and\ \citenamefont
  {d'Enterria}(2022)}]{Shao:2022cly}%
  \BibitemOpen
  \bibfield  {author} {\bibinfo {author} {\bibfnamefont {H.-S.}\ \bibnamefont
  {Shao}}\ and\ \bibinfo {author} {\bibfnamefont {D.}~\bibnamefont
  {d'Enterria}},\ }\bibfield  {title} {\bibinfo {title} {{gamma-UPC: automated
  generation of exclusive photon-photon processes in ultraperipheral proton and
  nuclear collisions with varying form factors}},\ }\href
  {https://doi.org/10.1007/JHEP09(2022)248} {\bibfield  {journal} {\bibinfo
  {journal} {JHEP}\ }\textbf {\bibinfo {volume} {09}},\ \bibinfo {pages}
  {248}},\ \Eprint {https://arxiv.org/abs/2207.03012} {arXiv:2207.03012
  [hep-ph]} \BibitemShut {NoStop}%
\bibitem [{\citenamefont {Coadou}(2022)}]{Coadou:2022nsh}%
  \BibitemOpen
  \bibfield  {author} {\bibinfo {author} {\bibfnamefont {Y.}~\bibnamefont
  {Coadou}},\ }\bibfield  {title} {\bibinfo {title} {{Boosted decision trees}}\
  }\href {https://doi.org/10.1142/9789811234033\_0002}
  {10.1142/9789811234033\_0002} (\bibinfo {year} {2022}),\ \Eprint
  {https://arxiv.org/abs/2206.09645} {arXiv:2206.09645 [physics.data-an]}
  \BibitemShut {NoStop}%
\bibitem [{\citenamefont {Cowan}(2021)}]{Cowan2021}%
  \BibitemOpen
  \bibfield  {author} {\bibinfo {author} {\bibfnamefont {G.}~\bibnamefont
  {Cowan}},\ }\bibinfo {title} {Statistics},\ in\ \href
  {https://doi.org/10.1007/978-3-319-93785-4_5} {\emph {\bibinfo {booktitle}
  {Handbook of Particle Detection and Imaging}}},\ \bibinfo {editor} {edited
  by\ \bibinfo {editor} {\bibfnamefont {I.}~\bibnamefont {Fleck}}, \bibinfo
  {editor} {\bibfnamefont {M.}~\bibnamefont {Titov}}, \bibinfo {editor}
  {\bibfnamefont {C.}~\bibnamefont {Grupen}},\ and\ \bibinfo {editor}
  {\bibfnamefont {I.}~\bibnamefont {Buvat}}}\ (\bibinfo  {publisher} {Springer
  International Publishing},\ \bibinfo {address} {Cham},\ \bibinfo {year}
  {2021})\ pp.\ \bibinfo {pages} {117--143}\BibitemShut {NoStop}%
\bibitem [{ATL(2021)}]{ATLAS:2021pce}%
  \BibitemOpen
  \bibfield  {title} {\bibinfo {title} {{Sensitivity to exclusive $WW$
  production in photon scattering at the High Luminosity LHC}},\ }\href@noop {}
  {\  (\bibinfo {year} {2021})}\BibitemShut {NoStop}%
\bibitem [{\citenamefont {Cepeda}\ \emph {et~al.}(2019)\citenamefont {Cepeda}
  \emph {et~al.}}]{Cepeda:2019klc}%
  \BibitemOpen
  \bibfield  {author} {\bibinfo {author} {\bibfnamefont {M.}~\bibnamefont
  {Cepeda}} \emph {et~al.},\ }\bibfield  {title} {\bibinfo {title} {{Report
  from Working Group 2}: {Higgs Physics at the HL-LHC and HE-LHC}},\ }\href
  {https://doi.org/10.23731/CYRM-2019-007.221} {\bibfield  {journal} {\bibinfo
  {journal} {CERN Yellow Rep. Monogr.}\ }\textbf {\bibinfo {volume} {7}},\
  \bibinfo {pages} {221} (\bibinfo {year} {2019})},\ \Eprint
  {https://arxiv.org/abs/1902.00134} {arXiv:1902.00134 [hep-ph]} \BibitemShut
  {NoStop}%
\bibitem [{\citenamefont {Arkani-Hamed}\ \emph {et~al.}(2016)\citenamefont
  {Arkani-Hamed}, \citenamefont {Han}, \citenamefont {Mangano},\ and\
  \citenamefont {Wang}}]{Arkani-Hamed:2015vfh}%
  \BibitemOpen
  \bibfield  {author} {\bibinfo {author} {\bibfnamefont {N.}~\bibnamefont
  {Arkani-Hamed}}, \bibinfo {author} {\bibfnamefont {T.}~\bibnamefont {Han}},
  \bibinfo {author} {\bibfnamefont {M.}~\bibnamefont {Mangano}},\ and\ \bibinfo
  {author} {\bibfnamefont {L.-T.}\ \bibnamefont {Wang}},\ }\bibfield  {title}
  {\bibinfo {title} {{Physics opportunities of a 100 TeV
  proton\textendash{}proton collider}},\ }\href
  {https://doi.org/10.1016/j.physrep.2016.07.004} {\bibfield  {journal}
  {\bibinfo  {journal} {Phys. Rept.}\ }\textbf {\bibinfo {volume} {652}},\
  \bibinfo {pages} {1} (\bibinfo {year} {2016})},\ \Eprint
  {https://arxiv.org/abs/1511.06495} {arXiv:1511.06495 [hep-ph]} \BibitemShut
  {NoStop}%
\bibitem [{\citenamefont {Klusek-Gawenda}\ \emph {et~al.}(2025)\citenamefont
  {Klusek-Gawenda}, \citenamefont {Goncalves},\ and\ \citenamefont
  {Szczurek}}]{Klusek-Gawenda:2025ffh}%
  \BibitemOpen
  \bibfield  {author} {\bibinfo {author} {\bibfnamefont {M.}~\bibnamefont
  {Klusek-Gawenda}}, \bibinfo {author} {\bibfnamefont {V.~P.}\ \bibnamefont
  {Goncalves}},\ and\ \bibinfo {author} {\bibfnamefont {A.}~\bibnamefont
  {Szczurek}},\ }\bibfield  {title} {\bibinfo {title} {{Light-by-Light
  scattering in ultraperipheral heavy ion collisions: Estimating inelastic
  contributions}},\ }\href {https://doi.org/10.1016/j.physletb.2025.139614}
  {\bibfield  {journal} {\bibinfo  {journal} {Phys. Lett. B}\ }\textbf
  {\bibinfo {volume} {867}},\ \bibinfo {pages} {139614} (\bibinfo {year}
  {2025})},\ \Eprint {https://arxiv.org/abs/2503.08624} {arXiv:2503.08624
  [hep-ph]} \BibitemShut {NoStop}%
\bibitem [{\citenamefont {D'Enterria}\ \emph {et~al.}(2019)\citenamefont
  {D'Enterria}, \citenamefont {Martins},\ and\ \citenamefont
  {Rebello~Teles}}]{DEnterria:2019cpw}%
  \BibitemOpen
  \bibfield  {author} {\bibinfo {author} {\bibfnamefont {D.}~\bibnamefont
  {D'Enterria}}, \bibinfo {author} {\bibfnamefont {D.}~\bibnamefont
  {Martins}},\ and\ \bibinfo {author} {\bibfnamefont {P.}~\bibnamefont
  {Rebello~Teles}},\ }\bibfield  {title} {\bibinfo {title} {{Two-photon fusion
  Higgs production in collisions with proton and ion beams at the LHC, HE-LHC,
  and FCC}},\ }\href@noop {} {\bibfield  {journal} {\bibinfo  {journal}
  {Frascati Phys. Ser.}\ }\textbf {\bibinfo {volume} {69}},\ \bibinfo {pages}
  {156} (\bibinfo {year} {2019})}\BibitemShut {NoStop}%
\end{thebibliography}%

\end{document}